\documentclass[prc,twocolumn,floatfix,showpacs,preprintnumbers,amsmath,amssymb]{revtex4}
\usepackage{graphicx}
\usepackage{epsfig}
\usepackage{CJK}
\usepackage{amsmath}
\usepackage{supertabular}
\usepackage{textcomp}
\usepackage{threeparttable}

\begin{document}


\title{Finite particle number description of neutron matter using the unitary correlation operator and high-momentum pair methods}
\author{Niu Wan,$^{1}$$\footnote{wanniu\underline{ }nju@163.com}$ Takayuki Myo,$^{2,3}$$\footnote{takayuki.myo@oit.ac.jp}$ Chang Xu,$^{1}$$\footnote{cxu@nju.edu.cn}$ Hiroshi Toki,$^{3}$$\footnote{toki@rcnp.osaka-u.ac.jp}$ Hisashi Horiuchi,$^{3}$$\footnote{horiuchi@rcnp.osaka-u.ac.jp}$ Mengjiao Lyu$^{4}$$\footnote{mengjiao.lyu@nuaa.edu.cn}$}
\address{$^{1}$School of Physics, Nanjing University, Nanjing 210093, China\\
$^{2}$General Education, Faculty of Engineering, Osaka Institute of Technology, Osaka, Osaka 535-8585, Japan\\
$^{3}$Research Center for Nuclear Physics (RCNP), Osaka University, Ibaraki, Osaka 567-0047, Japan\\
$^{4}$College of Science, Nanjing University of Aeronautics and Astronautics, Nanjing 210093, China }

\begin{abstract}
By using bare Argonne V4' (AV4'), V6' (AV6'), and V8' (AV8') nucleon-nucleon ($NN$) interactions respectively, the nuclear equations of state (EOSs) for neutron matter are calculated with the unitary correlation operator and high-momentum pair methods. The neutron matter is described under a finite particle number approach with magic number $N=66$ under a periodic boundary condition. The central short-range correlation coming from the short-range repulsion in the $NN$ interaction is treated by the unitary correlation operator method (UCOM) and the tensor correlation and spin-orbit effects are described by the two-particle two-hole (2p2h) excitations of nucleon pairs, in which the two nucleons with a large relative momentum are regarded as a high-momentum pair (HM). With the 2p2h configurations increasing, the total energy per particle of neutron matter is well converged under this UCOM+HM framework. By comparing the results calculated with AV4', AV6', and AV8' $NN$ interactions, the effects of the short-range correlation, the tensor correlation, and the spin-orbit coupling on the density dependence of the total energy per particle of neutron matter are demonstrated. Moreover, the contribution of each Hamiltonian component to the total energy per particle is discussed. The EOSs of neutron matter calculated within the present UCOM+HM framework agree with the calculations of six different microscopic many-body theories, especially in agreement with the auxiliary field diffusion Monte Carlo calculations.
\end{abstract}


\maketitle

\section{Introduction}
Properties of neutron matter play a crucial role in determining the structures of not only neutron stars \cite{JML,CJHo,AWS,BALi,BKSh,LWCh,DVS} but also extremely isospin-asymmetric nuclear system \cite{BSP,PDan,SGan,SHSh}. Though finite nuclei have provided a lot of information about nuclear matter at sub- and around saturation densities \cite{LWCh2,BGT,MAF,CXu2,CXu,NWa,Wan,Liu}, it is difficult to extrapolate the properties to higher-density region. In order to obtain reliable equation of state (EOS) of nuclear matter in the overall density region, a possible way is to perform microscopic many-body calculations based on bare nucleon-nucleon ($NN$) interactions. In short-distance region, the central force of the $NN$ interaction has a very strong repulsion core, while in intermediate- and long-distance region, there is a strong tensor force \cite{Wir1,Piep,Wir2}. The high-momentum components in nuclear system are mainly induced by above two kinds of forces. The short-range repulsion can decease the wave function amplitude of the relative motion at short-range distances for the two nucleons in a nucleon pair, while the tensor force can introduce the $D$-wave state because there can exist strong $S$-$D$ couplings.

Different approaches have been proposed to treat the $NN$ correlations coming from the $NN$ interaction, such as the utilization of correlation functions introducing the Jastrow factor and the renormalization of the $NN$ interaction using the unitary transformation. In our recent variational approach proposed for finite nuclei, namely the tensor-optimized antisymmetrized molecular dynamics (TOAMD) \cite{Myo11,Myo12,Myo13,Myo14,Myo15}, the central- and tensor-operator type correlation functions are employed to treat the short-range and tensor correlations, respectively. Within this method, the antisymmetrized-molecular-dynamics (AMD) \cite{Enyo} bases are used as basis wave functions. With bare $NN$ interactions, the above two correlations in $s$-shell nuclei have been successfully investigated as well as their structure and properties \cite{Myo12,Myo13,Myo14,Myo15}. Recently, Yamada proposed another variational theory to study the properties of nuclear matter, namely tensor-optimized Fermi sphere (TOFS) method \cite{Yama1,Yama2}. In the TOFS, the nuclear matter is described within a Fermi sphere, and the correlation functions multiplied to the Fermi-sphere state are also used to treat the $NN$ correlations in nuclear matter. The minimal energy of nuclear matter is searched in order to determine the parameters in the functions. With Argonne V4' (AV4') central $NN$ potential \cite{Wir2} which can induce the short-range correlation, the obtained EOS of symmetric nuclear matter within TOFS agrees with the benchmark results calculated with other theories \cite{Yama1,Yama2}.

The short-range correlation can be also treated by the unitary correlation operator method (UCOM) by employing the unitary correlation operator \cite{ucom1,ucom2,ucom3}. Due to the short-range repulsion, the wave function amplitudes for the two nucleons in a nucleon pair are decreased for the relative motion at short-range distances. It has been shown that the UCOM can describe very well the short-range correlation in finite nuclei \cite{ucom1,ucom2,ucom3,Myo1,Myo2}. Under the relativistic and nonrelativistic frameworks within Hartree-Fock (HF) approximation for Fermi sphere \cite{Hu1,Hu2}, Hu \emph{et al.} employed the UCOM to further describe the short-range correlation in neutron matter. The obtained EOS of neutron matter agrees very well with those calculated by the Relativistic Brueckner-Hartree-Fock theory, which indicates that the short-range correlation in neutron matter can also be treated by the UCOM \cite{Hu1,Hu2}.

In order to describe the tensor correlation, some approaches different from employing correlation functions have also been successfully proposed in our previous works for finite nuclei. As is well known that the two-particle two-hole (2p2h) excitations can describe the strong tensor correlation coming from the tensor force in nucleon pairs, the tensor correlation can be described by optimizing the 2p2h configurations, as introduced in the tensor-optimized shell model (TOSM) \cite{ucom3,Myo1,Myo2}. In the TOSM, the total wave function is superposed by the standard shell-model state and sufficient 2p2h states. By optimizing the 2p2h configurations without truncation of the particle states, the tensor correlation can be treated based on the framework of the shell model. Within the extension of HF theory \cite{Hu3}, the 2p2h configurations are employed to describe the high-momentum components of nuclear matter. The obtained results are similar to those of Brueckner-Hartree-Fock (BHF) theory and the corresponding momentum distribution is found to have high-momentum components due to the 2p2h excitations. In our recently developed approach, namely high-momentum AMD (HM-AMD) \cite{Myo3,Lyu1,Lyu2,Myo4,Zhao}, the high-momentum nucleon pair (HM) is introduced, in which the two nucleons for a 2p2h excitation both involve a large transfer momentum with an opposite direction \cite{Itag}. Similar to the TOAMD approach, the AMD bases \cite{Enyo} are used as basis wave functions. However, the 2p2h excitations are employed in HMAMD to treat the $NN$ correlations instead of the correlation functions used in TOAMD. This new approach has described very well the high-momentum components in finite nuclei \cite{Myo3,Lyu1,Lyu2,Myo4,Zhao}.

In the study of nuclear matter, there are several other microscopic theories treating the $NN$ correlations starting from the bare $NN$ interactions, such as BHF and Brueckner-Bethe-Goldstone (BBG) \cite{BHF1,BHF2,BHF3}, self-consistent Green's function (SCGF) \cite{SCGF1,SCGF2,SCGF3,SCGF4}, Fermi hypernetted chain (FHNC) \cite{FHNC1,FHNC2,FHNC3,FHNC4}, auxiliary field diffusion Monte Carlo (AFDMC) \cite{AFDMC1,AFDMC2,AFDMC3}, Green's function Monte Carlo (GFMC) \cite{GFMC1,GFMC2}, and coupled cluster theory (CC) \cite{CC1,CC2,CC3}. The BHF approach can be interpreted as the lowest order under the framework of the BBG theory. The ground-state energy of the latter is calculated by employing the linked cluster expansion by means of the $G$-matrix, which is ordered based on the number of independent hole lines. The diagram corresponding to the hole line with a number $n$ describes the $n$-body correlations. In the BHF approach, the total energy is calculated within the truncation of two hole-line approximation, which only includes the two-body correlations. In the SCGF approach, the total energy is calculated from the in-medium one-body propagator, which is obtained from the Dyson equation by using the ladder diagram expansion. The FHNC approach is one of the variational methods different from above nonperturbative ones. With a given trial wave function multiplied by correlation operators, the total energy can be evaluated within the cluster expansion by using the FHNC integral equations \cite{FHNC1}. Besides, both the AFDMC and the GFMC approaches are extended for nuclear systems under the framework of Quantum Monte Carlo (QMC) method, which has successfully described the ground state of infinite atomic systems \cite{QMC1,QMC2}. The difference between AFDMC and GFMC mainly exists in their treatments about the spin and isospin channels. The AFDMC samples the spin-isospin states according to the Hubbard-Stratonovich auxiliary fields, while the GFMC sums all the states. Compared with AFDMC, GFMC can treat nuclear systems with more accuracy but smaller mass number. By comparing the benchmark results calculated by Baldo \emph{et al.} under above different theories for nuclear matter with several kinds of the Argonne bare $NN$ interactions \cite{BHF1}, very similar density dependence of the total energy per particle of nuclear matter are obtained. In detail, if only the central potential including the short-range repulsion is considered, the calculated results agree with each other very well. However, when the tensor and spin-orbit forces are additionally included in the $NN$ potentials, there will be large differences among the density dependences especially for the symmetric nuclear matter \cite{BHF1}, which mainly come from the different treatments of the tensor force as tensor correlation. The intermediate- and long-range properties of the tensor correlation coming from the $NN$ interaction can induce many-body correlations, which is very important for nuclear matter studies. In particular, the saturation property of symmetric nuclear matter has a very close relation to the tensor correlation. Besides of the bare two-body $NN$ force, there are also studies about the effects of many-body forces on the properties of nuclear matter, especially three-body force (TBF) \cite{TBF0,TBF1,TBF2,TBF3,TBF4,TBF5}. According to the detailed analysis with several different forms for TBF in Ref. \cite{TBF4}, it is found that the TBF can have a large impact on the EOS of nuclear matter at high densities. Since there exists large uncertainty in many-body forces at present, we mainly concentrate on the bare two-body $NN$ interaction to investigate the properties of neutron matter in this study.

In our very recent work, we have proposed a new variational approach for nuclear matter description starting from the bare $NN$ interaction \cite{ucom4}. Under a periodic boundary condition, the nuclear matter is described in a cubic box with finite size. This finite particle number approach has been successfully used in AFDMC, GFMC, CC, and electron systems \cite{Lin}. Allowing for the different correlations coming from the $NN$ interaction, the UCOM is employed in the new variational method to treat the central short-range correlation induced by the short-range repulsion. In addition, the 2p2h excitations of nucleon pairs are further included to describe the high-momentum components in nuclear matter. The 2p2h configurations are added into the total wave function in the same manner for finite nuclei \cite{ucom3,Myo1,Myo2}. The two nucleons in the nucleon pair for a 2p2h excitation both obtain a transfer momentum with opposite directions, which results in a large relative momentum for the two nucleons of a 2p2h excitation. The above new framework for nuclear matter is named UCOM+HM \cite{ucom4}.

We have investigated the validity of the present UCOM+HM framework to study the neutron and symmetric nuclear matters in the previous work \cite{ucom4}. From the calculation for nuclear matter with AV4' central potential involving the short-range repulsion, we confirm the applicability of the UCOM to treat the central short-range correlation. Besides, the additional 2p2h excitations are found to contribute to the total energy around the normal nuclear density in nuclear matter by several MeV per particle. The EOSs provided by the present method for the neutron and symmetric nuclear matters are consistent to those with other theories from low-density to high-density regions \cite{ucom4}, which indicates the reliability of this new variational approach.

In present work, we concentrate on the effects of the tensor correlation and spin-orbit coupling on the properties of neutron matter. The tensor correlation is well known to be very important for the symmetric nuclear matter, especially the saturation property. It is also very interesting to investigate the tensor correlation in neutron matter. With Argonne V6' (AV6') and V8' (AV8') $NN$ interactions \cite{Wir2}, the effects of not only the tensor correlation but also the spin-orbit coupling on the EOS of neutron matter are studied. Their respective contributions to the total energy of neutron matter are obtained as well. The corresponding EOSs of neutron matter calculated with different $NN$ interactions are compared with those of several other microscopic many-body theories as well. In Section \ref{ii2}, the detailed formulism of this new variational framework is given. In Section \ref{iii3}, the calculated results of neutron matter are presented. A summary is given in Section \ref{iv4}.

\section{Formulism}
\label{ii2}
\subsection{Bare nucleon-nucleon ($NN$) interaction}
One of the most accurate bare nucleon-nucleon ($NN$) interactions currently is the Argonne V18 (AV18) potential \cite{Wir1,Piep,Wir2}. Because of its sophisticated operatorial structure for some many-body schemes, several simplified versions of AV18 potential are devised for benchmark purposes, such as AV4', AV6', and AV8' $NN$ interactions \cite{Wir2}. The AV8' potential can be written as the summation of the first eight components of AV18 potential by
\begin{align}\label{eqiia1}
V_{ij}=\sum_{k=1,...,8}v_k(r_{ij})O^k_{ij}.
\end{align}
The other six quadratic spin-orbit and four charge-dependent components of AV18 potential are removed for AV8' version, and the radial functions $v_k(r_{ij})$ are readjusted to preserve experimental data on phase shifts and properties of deuteron. The eight operators $O^k_{ij}$ in Eq. (\ref{eqiia1}) include the spin, isospin, tensor, spin-orbit coupling components of the nuclear force \cite{Wir2}:
\begin{align}\label{eqiia2}
O^{k=1,...,8}_{ij}=
&1,\ \boldsymbol{\sigma}_i\cdot\boldsymbol{\sigma}_j,\ \boldsymbol{\tau}_i\cdot\boldsymbol{\tau}_j,\ (\boldsymbol{\sigma}_i\cdot\boldsymbol{\sigma}_j)(\boldsymbol{\tau}_i\cdot\boldsymbol{\tau}_j),\nonumber\\
&S_{ij},\ S_{ij}(\boldsymbol{\tau}_i\cdot\boldsymbol{\tau}_j),\nonumber\\
&\boldsymbol{L}\cdot\boldsymbol{S},\ \boldsymbol{L}\cdot\boldsymbol{S}(\boldsymbol{\tau}_i\cdot\boldsymbol{\tau}_j).
\end{align}
By removing the last two spin-orbit coupling components and refitting the radial functions, the first six components in Eq. (\ref{eqiia2}) constitute the AV6' potential. Similarly, without considering the tensor and spin-orbit coupling components, the AV4' central potential only includes the first four components of Eq. (\ref{eqiia2}), but the radial functions have been refitted by the deuteron binding energy. With these simplified $NN$ interactions, namely AV4', AV6', and AV8', the EOSs of neutron matter will be microscopically calculated in present work and the effects of the tensor correlation and spin-orbit coupling on the results will be discussed as well.

\subsection{Wave function}
As introduced in our recent work \cite{ucom4}, the 0p0h state of neutron matter is defined by the Slater-determinant as
\begin{align}\label{eqiib1}
|\textmd{0p0h}\rangle
&=\frac{1}{\sqrt{N!}}\textmd{det}\left\{\prod^N_{i=1}\phi_{\alpha_i}(\boldsymbol{r}_i)\right\},\\\label{eqiib2}
\phi_{\alpha}(\boldsymbol{r})
&=\frac{1}{\sqrt{L^3}}e^{i\boldsymbol{k}_{\alpha}\cdot\boldsymbol{r}}\chi^{\sigma}_{\alpha},\\\label{eqiib3}
\langle\phi_{\alpha}|\phi_{\alpha'}\rangle
&=\delta_{\alpha,\alpha'},
\end{align}
where $N$ is the neutron number of neutron matter, $\phi_{\alpha}(\boldsymbol{r})$ is the plane wave function of a neutron $\boldsymbol{k}_{\alpha}$ is the momentum, and $\chi^{\sigma}_{\alpha}$ is the spin component which can be up or down. The index $\alpha$ represents the quantum number for both momentum and spin. The neutron matter is described in a cubic box with finite size $L$ as given in Eq. (\ref{eqiib2}), which is determined by the wave function normalization. With the periodic boundary condition $\phi_{\alpha}(\boldsymbol{r}+L\hat{\boldsymbol{x}})=\phi_{\alpha}(\boldsymbol{r})$ employed in the description of neutron matter, the neutron momentum is discretized by the gap $\Delta k=\frac{2\pi}{L}$. By using an integer vector $\boldsymbol{n}=(n_x, n_y, n_z)$, the momentum of each neutron can be calculated by $\boldsymbol{k}=\frac{2\pi}{L}\boldsymbol{n}$. Within a lattice in momentum space corresponding to the cubic box in coordinate space, as shown in our previous work \cite{ucom4}, the momentum eigenstate of each neutron can be represented by the grid point.

On account of the periodic boundary condition and the symmetry of the wave function in Eq. (\ref{eqiib1}), there exist magic particle numbers corresponding to the shell closures, where the corresponding grid points are $N_g=1, 7, 19, 27, 33, 57, ...$. For each grid number $N_g$, besides of the single neutron wave function with momentum $\boldsymbol{k}_i$ ($i=1, ..., N_g$) in 0p0h state, the total particle number becomes $N=2N_g$ due to the spin.

For the 2p2h excitations of neutron matter, the 2p2h configurations can be written as
\begin{align}\label{eqiib4}
|\textmd{2p2h}\rangle=|mn;i^{-1}j^{-1}\rangle=a^{\dag}_ma^{\dag}_na_ia_j|\textmd{0p0h}\rangle,
\end{align}
where the index $i$ and $j$ ($i, j=1, ..., N$) represent hole states from lower magnitude of momenta, and the index $m$ and $n$ ($m, n>N$) are particle states in 2p2h configurations. The total momentum between two holes and two particles are conserved with the following condition:
\begin{align}\label{eqiib5}
&\boldsymbol{k}_i+\boldsymbol{k}_j=\boldsymbol{k}_m+\boldsymbol{k}_n,\\\label{eqiib6}
&\boldsymbol{k}_m=\boldsymbol{k}_i+\boldsymbol{q},\quad
\boldsymbol{k}_n=\boldsymbol{k}_j-\boldsymbol{q}.
\end{align}
The quantity $\boldsymbol{q}=\frac{2\pi}{L}\boldsymbol{n}_q$ is the transfer momentum to link the two particles and two holes in a 2p2h configuration, which is correlated with the relative momentum of the two particle states. Besides, due to the periodic boundary condition, the transfer momentum is also discretized with the mode $\boldsymbol{n}_q=(n_{qx}, n_{qy}, n_{qz})$. If the transfer momentum is large enough, the high-momentum components in neutron matter can be naturally induced by the 2p2h excitations. For example, if the neutron number $N=66$ and the normal density $\rho=0.17~\textmd{fm}^{-3}$ are taken, the magnitude of the transfer momentum is about $|\boldsymbol{q}|=\frac{2\pi}{L}|\boldsymbol{n}_q|=5.2~\textmd{fm}^{-1}$ with the integer mode $|\boldsymbol{n}_q|=6$. Compared with the empirical Fermi momentum $k_F=1.4~\textmd{fm}^{-1}$, it is large enough to excite nucleons to high-momentum regions. The high-momentum component is closely related to the descriptions of both the short-range and the tensor correlations \cite{Lyu1,Lyu2,Myo4,Zhao}. In present study, a maximum integer for the magnitude of the transfer mode $\boldsymbol{n}_q$ as $n^{\textmd{max}}_q$ is employed with the condition $n^{\textmd{max}}_q\geqslant|\boldsymbol{n}_q|$. The number of the total 2p2h configurations is determined by the quantity $n^{\textmd{max}}_q$, which affects the basis space for present calculation. By increasing this parameter, we first check the energy convergence of neutron matter at normal density.

\begin{figure}[thb]
\centering
\includegraphics[width=0.95\linewidth]{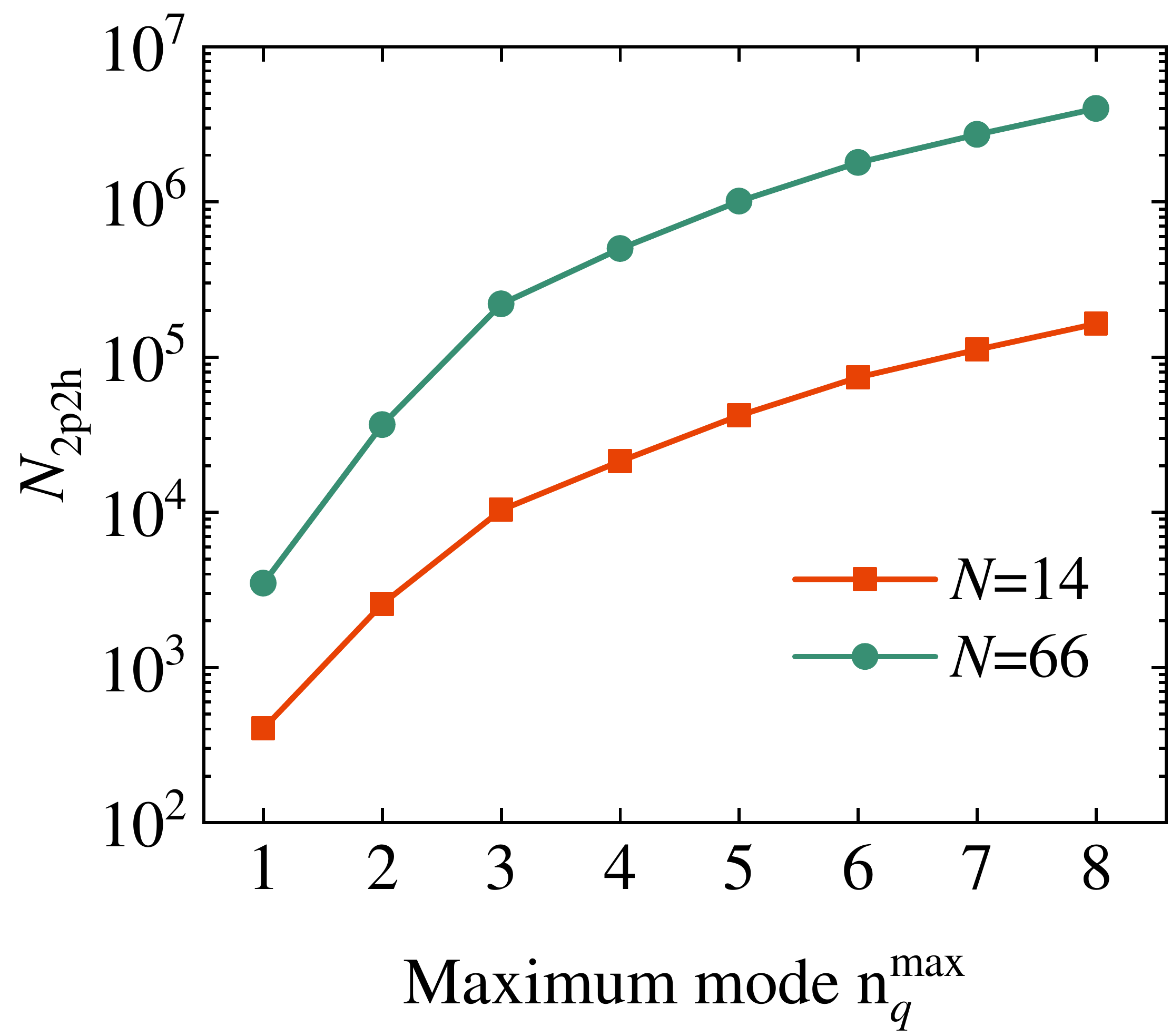}
\caption{Number of the total 2p2h configurations $N_{\textmd{2p2h}}$ varied with the maximum mode $n^{\textmd{max}}_q$ of the transfer momentum in neutron matter by taking neutron magic numbers $N=14$ and $N=66$ as examples.}
\label{figNum2p2h}
\end{figure}

By superposing the 0p0h and 2p2h configurations, the total wave function $\Phi$ of neutron matter can be written as \cite{ucom4}:
\begin{align}\label{eqiib8}
\Phi
&=C_0|\textmd{0p0h}\rangle+\sum^{N_{\textmd{2p2h}}}_{p=1}C_p|\textmd{2p2h},p\rangle,
\end{align}
where $N_{\textmd{2p2h}}$ is the number of the total 2p2h configurations and $\{C_p\}$ are the configuration amplitudes, which can be variationally determined. The index $p$ represents each configuration, where $p=0$ corresponds to the 0p0h state.
In Fig. \ref{figNum2p2h}, by taking the neutron magic numbers $N=14$ and $N=66$ as examples, we show the variation of the number $N_{\textmd{2p2h}}$ with the maximum transfer mode $n^{\textmd{max}}_q$. As shown in Fig. \ref{figNum2p2h}, the number of the total 2p2h configurations approximately exponentially increases with the maximum transfer mode. That is to say, the mode space of the present calculation will increase sharply with the maximum transfer mode.

\subsection{Unitary correlation operator method (UCOM)}
For the short-range repulsion in $NN$ interaction, the UCOM is further used to treat the corresponding central short-range correlation in neutron matter. Within UCOM, the correlated wave function $\Psi$ considering the short-range correlation in the nuclear system can be written by multiplying a unitary correlation operator $C_r$ to the uncorrelated one $\Phi$: $\Psi=C_r\Phi$, where the unitary correlation operator $C_r$ is defined as \cite{ucom1,ucom2}
\begin{align}\label{eqii9}
C_r=\exp\left(-i\sum_{i<j}g_{ij}\right)=\prod^N_{i<j}c_{r,ij},
\end{align}
where $g$ is a pair-type Hermite generator and $c_{r,ij}$ is for a nucleon pair. The specific form for the operator $g$ can be written as
\begin{align}\label{eqii10}
g=\frac{1}{2}\{p_rs(r)+s(r)p_r\},
\end{align}
where $p_r$ is the parallel relative momentum between nucleons and $s(r)$ is the variation of the relative wave function. By using the transformation $\Psi=C_r\Phi$ under UCOM, the transformed Schr\"{o}dinger equation for $\Phi$ can be written as $\tilde{H}\Phi=E\Phi$, where $\tilde{H}$ is the transformed Hamiltonian from the original one $H=T+V$ as
\begin{align}\label{eqii13}
\tilde{H}&=C^{\dag}_rHC_r=C^{\dag}_rTC_r+C^{\dag}_rVC_r=\tilde{T}+\tilde{V},\\\label{eqii14}
\tilde{T}&=\sum^N_{i=1}t_i+\sum^N_{i<j}u_{ij},\quad \tilde{V}=\sum^N_{i<j}\tilde{v}_{ij}.
\end{align}
With the unitary correlation operator $C_r$, we first derive the transformed Hamiltonian $\tilde{H}$ within the UCOM. Then by using the total wave function $\Phi$ given in Eq. (\ref{eqiib8}), the Hamiltonian matrix $\mathcal {H}$ can be calculated for different Hamiltonian components as
\begin{align}\label{eqad10}
\mathcal {H}=
\left(
\begin{array}{ccc}
\langle0|\tilde{H}|0\rangle     & \langle0|\tilde{H}|2,p=1\rangle     & ... \\
\langle2,p'=1|\tilde{H}|0\rangle & \langle2,p'=1|\tilde{H}|2,p=1\rangle & ... \\
...                             & ...                                 & ...
\end{array}
\right),
\end{align}
where $|0\rangle$ and $|2,p\rangle$ denote the configurations $|\textmd{0p0h}\rangle$ and $|\textmd{2p2h},p\rangle$, respectively. Within the power method \cite{Power1,Power2,Power3}, we can solve the energy eigenvalue problem for the Hamiltonian matrix with $N_{\textmd{2p2h}}+1$ dimension. Then the configuration amplitudes $\{C_p\}$ in Eq. (\ref{eqiib8}) can be variationally determined by minimizing the total energy as well as each Hamiltonian component energy. Actually, from Eq. (\ref{eqii9}) we can see that since the operator $C_r$ is a many-body operator, we also confront a many-body problem about the transformed Hamiltonian $\tilde{H}$. However, it is reasonable to take the two-body operator truncation for the short-range correlation case, as discussed in previous studies \cite{ucom1,ucom2,ucom3,Myo1,Myo2}. For the transformed kinetic part $\tilde{T}$ in Eq. (\ref{eqii14}), there are two terms: the uncorrelated one-body term $t_i$ and the correlated two-body term $u_{ij}$, where the latter comes from the short-range correlation between nucleons and is closely related to both the momentum and angular momentum of the relative motion:
\begin{align}\label{eqiiad1}
u(r)=w(r)+\frac{1}{2}\left[p^2_r\frac{1}{2\mu_r(r)}+\frac{1}{2\mu_r(r)}p^2_r\right]+\frac{\boldsymbol{L}^2}{2\mu_{\Omega}(r)r^2},
\end{align}
where the forms for the functions $w(r)$, $\mu_r(r)$, and $\mu_{\Omega}(r)$ are respectively \cite{ucom2}
\begin{align}\label{eqiiad2}
w(r)&=\frac{\hbar^2}{m}\left(\frac{7}{4}\frac{R^{''2}_{+}(r)}{R^{'4}_{+}(r)}-\frac{1}{2}\frac{R^{'''}_{+}(r)}{R^{'3}_{+}(r)}\right),\\
\frac{1}{2\mu_r(r)}&=\frac{1}{m}\left(\frac{1}{R^{'2}_{+}(r)}-1\right),\\
\frac{1}{2\mu_{\Omega}(r)}&=\frac{1}{m}\left(\frac{r^2}{R^{2}_{+}(r)}-1\right),
\end{align}
where the function $R_+(r)$ is usually employed to replace the previous function $s(r)$ for UCOM calculations. For the two-body potential part $\tilde{V}$ in Eq. (\ref{eqii14}), the potential energy $\tilde{v}$ is also related to the function $R_+(r)$ and is usually calculated by $v(R_+(r))$. The function $R_+(r)$ stands for the transformed relative distance from the original one $r$ \cite{ucom1,ucom2}. The relation between the two functions $R_+(r)$ and $s(r)$ is
\begin{align}\label{eqii11}
\frac{dR_+(r)}{dr}&=\frac{s[R_+(r)]}{s(r)},\\\label{eqii12}
c^{\dag}_rrc_r&=R_+(r).
\end{align}
The function $R_+(r)$ can decease the wave function amplitude of the relative motion of the two nucleons in a nucleon pair at short-range distances, which can simulate the effect of the short-range correlation very well. The specific forms of $R_+(r)$ for the even (odd) channel with positive (negative) parity are respectively given as \cite{ucom1,ucom2}
\begin{align}\label{eqii15}
R^{\textmd{even}}_+(r)&=r+\alpha\left(\frac{r}{\beta}\right)^{\gamma}\exp\left[-\exp\left(\frac{r}{\beta}\right)\right],\\
R^{\textmd{odd}}_+(r)&=r+\alpha\left[1-\exp\left(-\frac{r}{\delta}\right)\right]\exp\left[-\exp\left(\frac{r}{\beta}\right)\right],
\end{align}
where $\alpha$, $\beta$, $\gamma$, and $\delta$ are parameters variationally determined by minimizing the total energy per particle of the 0p0h state within UCOM for neutron matter, which corresponds to 0p0h+UCOM calculation. The values of the parameters $\alpha$, $\beta$, $\gamma$, and $\delta$ are listed in Table \ref{ta01}. For neutron matter, the parameters of $R_+(r)$ are naturally determined to be the same for AV4', AV6', and AV8' $NN$ interactions. That is because for above three $NN$ interactions, the differences among them exist in the isospin-singlet channels $^3E$ and $^1O$, which are absent in neutron matter. For the two active channels $^1E$ and $^3O$ in neutron matter, the corresponding central parts are identical. In addition, the tensor and spin-orbit forces are not included in the process of determining the values of the three parameters, i.e., only the 0p0h+UCOM calculation involved. Since the number of the 2p2h configurations increases very sharply with the increasing maximum transfer mode, as shown in Fig. \ref{figNum2p2h}, it is very difficult to determine the parameters with UCOM+HM calculation. Besides, we find that the small changes of the parameters slightly affect the total energy. Hence, we determine the values of the parameters at 0p0h+UCOM level without much loss of accuracy.

\begin{table}[thb]
\centering
\caption{Values of the parameters $\alpha$, $\beta$, $\gamma$, and $\delta$ in the function $R_+(r)$ for UCOM with $^1E$ and $^3O$ channels in neutron matter.}
\label{ta01}
\begin{tabular}{ccccc}
\hline\hline
\vspace{-4mm}
\kern10mm &\kern17mm &\kern17mm &\kern17mm &\kern17mm\\
       & $\alpha$~[\textrm{fm}] & $\beta$~[\textrm{fm}] & $\gamma$ & $\delta$~[\textrm{fm}]     \\\hline
$^1E$  &      1.33       &      1.00      &      0.31      &                 \\
$^3O$  &      0.65       &      1.39      &                &      0.24       \\\hline\hline
\end{tabular}%
\end{table}

For the UCOM transformation, there have been two-body correlations contained in the wave function. In addition, as shown in Eq. (\ref{eqiib8}), the 2p2h configurations with high-momentum components are also included in the total wave function. Physically, there are up to 4p4h correlations involved in present approach. This new method is named as UCOM+HM hereafter.

\section{Numerical results}
\label{iii3}

\begin{figure}[thb]
\centering
\includegraphics[width=1.0\linewidth]{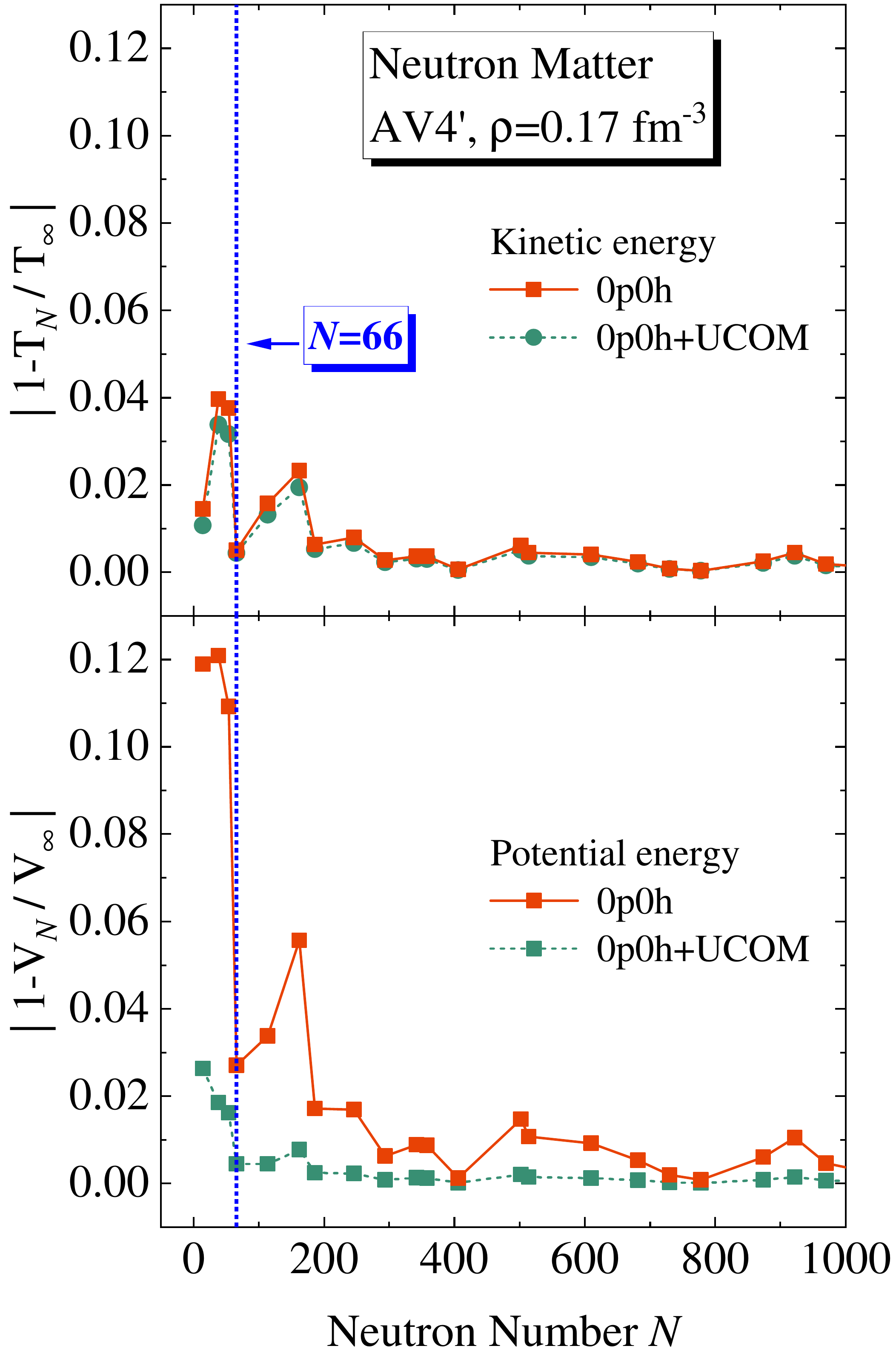}
\caption{Relative errors of the Hamiltonian component energy for both 0p0h and 0p0h+UCOM wave functions calculated with AV4' potential.}
\label{fig66}
\end{figure}

At first, we check the validity of the present finite particle number approach for neutron matter. We first calculate the HF energies for infinite neutron matter and the energies of finite neutron matter system with the finite particle number approach. In Fig. \ref{fig66}, we show the absolute values of the relative error of the Hamiltonian component energy changing with the neutron magic number for both 0p0h and 0p0h+UCOM wave functions. The calculations with AV4' potential and normal density $\rho=0.17~\textmd{fm}^{-3}$ are demonstrated as an example. The quantities $T_{N}$ and $V_{N}$ are the kinetic and potential energies for finite neutron matter system, while $T_{\infty}$ and $V_{\infty}$ are those for infinite neutron matter which can be directly calculated by using the Fermi sphere wave function with the infinite-space integral from the Hamiltonian matrix elements.

As shown in Fig. \ref{fig66}, the relative errors for both the kinetic and potential energies roughly decrease with the increasing neutron number $N$. In small neutron number region, no matter for the kinetic energy or for the potential energy, the relative energy errors between the finite particle number approach and the infinite neutron matter are the smallest with neutron magic number $N=2N_g=66$. In our recent work \cite{ucom4}, the relationship between the total energy of nuclear matter and the particle magic numbers has been investigated. The calculated results confirm that with the grid number $N_g=33$, i.e., the neutron magic number $N=66$, the numerical results of the kinetic and potential energies per particle for 0p0h state within the present approach can provide the best simulation for those of the infinite nuclear matter in the HF level for smaller magic particle numbers \cite{ucom4}. The same conclusion has also been obtained by other microscopic calculations for nuclear matter with this finite particle number approach, such as AFDMC \cite{AFDMC1,AFDMC2,AFDMC3}, GFMC \cite{GFMC1}, and CC \cite{CC2}. As a result, the neutron number $N=2N_g=66$ is chosen throughout the calculations in present study for neutron matter.

Besides, it can also be seen from Fig. \ref{fig66} that for the kinetic energy the difference of the relative error between 0p0h and 0p0h+UCOM calculations is relatively small. With neutron magic number $N=66$, the values are about 0.5\% and 0.4\% for 0p0h and 0p0h+UCOM cases, respectively. However, the relative errors for the potential energy calculated with 0p0h+UCOM wave function are generally much smaller than those with 0p0h wave function, especially in small neutron number region. With neutron magic number $N=66$, the relative error decreases from 2.7\% for 0p0h case to 0.4\% for 0p0h+UCOM case. This indicates that the UCOM can well treat the short-range correlation coming from the $NN$ interaction in neutron matter.

\begin{figure*}[thb]
\centering
\includegraphics[width=0.6\linewidth]{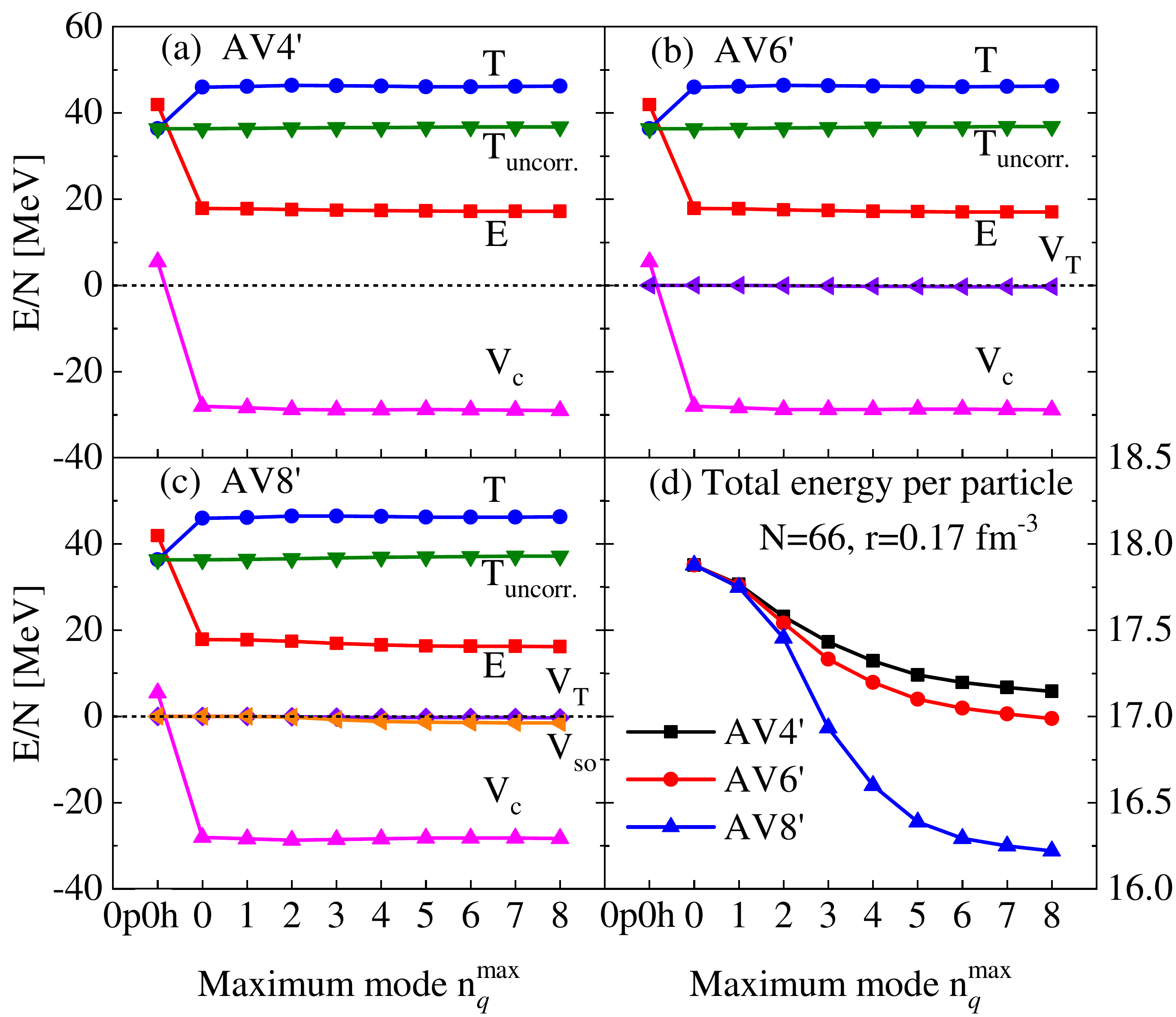}
\caption{Convergence of the Hamiltonian components as well as the total energy per particle for neutron matter with the increasing maximum mode of transfer momentum $n^{\textmd{max}}_q$ in UCOM+HM. The energies at $n_q=0$ correspond to the 0p0h+UCOM calculation. The panels (a), (b), and (c) correspond to the results calculated with AV4', AV6', and AV8' $NN$ interactions, respectively. The term $E$ is the total energy, $T$ is the total kinetic energy, $T_{\textmd{uncorr.}}$ is the uncorrelated one-body kinetic energy, and $V_{\textmd{c}}$, $V_{\textmd{T}}$, and $V_{\textmd{so}}$ are the potential energies corresponding to the central, tensor, and spin-orbit forces of the $NN$ interactions, respectively. The total energy per particle $E$ converged with the maximum mode of transfer momentum $n^{\textmd{max}}_q$ is shown in panel (d) for above three potentials.}
\label{fignq}
\end{figure*}

As mentioned before, the maximum mode of transfer momentum $n^{\textmd{max}}_q$ is introduced to control the model space of present calculation. So we first check whether the total energy per particle can converge with the increasing value of the integer mode $n^{\textmd{max}}_q$. In Fig. \ref{fignq} we show the dependences of the energies per particle on the maximum mode $n^{\textmd{max}}_q$ within the UCOM+HM by including 2p2h excitations to describe the high-momentum components in neutron matter. The mass number of neutron matter is set to be $N=66$ and the neutron density is the normal value $\rho=0.17~\textmd{fm}^{-3}$.

The Hamiltonian components per particle varied with the maximum mode $n^{\textmd{max}}_q$ corresponding to AV4', AV6', and AV8' $NN$ interactions are shown in the panels (a), (b), and (c) of Fig. \ref{fignq}, respectively. The energies at $n_q=0$ correspond to the 0p0h+UCOM calculation. It is clearly shown in our previous work \cite{ucom4} that the finite particle number approach with 0p0h and 0p0h+UCOM wave functions by using the neutron number $N=66$ can reproduce very well the EOSs with Fermi sphere in HF and HF+UCOM levels for neutron matter, respectively, which confirms the validity of the present calculations. By comparing the results calculated with 0p0h and 0p0h+UCOM wave functions, it can be seen in Fig. \ref{fignq} that for each $NN$ interaction case, there exists a large rise for the kinetic energy and a larger fall for the central potential, leading to a smaller total energy. This is because of the short-range correlation induced by the short-range repulsion in the central force, which has been successfully treated by UCOM. Due to the amplitude decrease of nucleon pairs induced by the short-range correlation at short-range distances, there are more attractive central force and additional correlated two-body part of the kinetic energy as shown in Eq. (\ref{eqii14}). We can obtain the effect of the short-range correlation from the correlated two-body part of the kinetic energy as a results of the difference between the total one $T$ and the uncorrelated one $T_{\textmd{uncorr.}}$, where the latter corresponds the one-body operator $\sum^N_{i=1}t_i$ in Eq. (\ref{eqii14}) without UCOM. The contributions of the short-range correlation to the kinetic energy for neutron matter all amount to about 9 MeV per particle for AV4', AV6', and AV8' potentials. Furthermore, with 0p0h+UCOM+HM wave function by including the 2p2h configurations, the convergences of the Hamiltonian components per particle for AV4', AV6', and AV8' potentials are clearly confirmed for every component, as shown in Fig. \ref{fignq}. The very similar behavior for above three potentials is also obtained that under the consideration of the 2p2h excitations as high-momentum pairs, the kinetic energy slightly increases as well as the potential energy.

The variation of the total energy per particle with the maximum mode of transfer momentum $n^{\textmd{max}}_q$ for above three $NN$ interactions is shown in the panel (d) of Fig. \ref{fignq}. It is apparent in the panel (d) that the total energies per particle converge well with the increasing maximum mode $n^{\textmd{max}}_q$ for the three potentials. Though the number of the 2p2h configurations $N_{\textmd{2p2h}}$ increases more than two times from $1.8\times10^6$ with $n^{\textmd{max}}_q=6$ to $4.0\times10^6$ with $n^{\textmd{max}}_q=8$, the energy differences between the two maximum modes for the three potentials are only tens of keV. Compared with the values calculated with $n^{\textmd{max}}_q=6$, the energies with $n^{\textmd{max}}_q=8$ only decrease about 0.3\%, 0.3\%, and 0.4\% for AV4', AV6', and AV8' $NN$ interactions, respectively. Allowing for the very sharp increase of the model space and the very small energy decrease, it is reasonable to take the maximum mode of transfer momentum $n^{\textmd{max}}_q=6$ to perform all the following calculations.

\begin{table*}[thb]
\centering
\caption{Values of the total energy per particle at the normal density $\rho=0.17~\textmd{fm}^{-3}$ calculated with 0p0h, 0p0h+UCOM, and UCOM+HM wave functions for AV4', AV6', and AV8' $NN$ interactions, respectively. The results for neutron number $N=14$ and $N=66$ cases are both given here. The quantity $\Delta_{\textmd{UCOM}}$ denotes the difference of the total energy calculated by 0p0h and 0p0h+UCOM wave functions, which corresponds to the contribution of the short-range correlation. Similarly, the one $\Delta_{\textmd{HM}}$ denotes the difference of the total energy calculated by 0p0h+UCOM and 0p0h+UCOM+HM (with $n^{\textmd{max}}_q=6$) wave functions, which represents the contribution of the high-momentum pairs.}
\label{ta03}
\begin{tabular}{ccccccccc}
\hline\hline
\vspace{-4mm}
\kern20mm  &\kern18mm  &\kern20mm  &\kern20mm     &\kern20mm  &\kern1mm  &\kern20mm  &\kern20mm  &\kern20mm  \\
                         &   $n^{\textmd{max}}_q$ &  \multicolumn{3}{c}{$E(N=14)$\ (MeV)} &   & \multicolumn{3}{c}{$E(N=66)$\ (MeV)} \\\cline{3-5}\cline{7-9}
                         &                          &  AV4'   &  AV6'    & AV8'     & &    AV4'  &  AV6'    & AV8'     \\\hline
0p0h                     &                          & 42.064  &  42.064  &  42.064  & &  41.873  &  41.873  &  41.873  \\\hline
0p0h+UCOM                &         0                & 17.963  &  17.963  &  17.963  & &  17.878  &  17.878  &  17.878  \\\hline
                         &         1                & 17.629  &  17.505  &  17.311  & &  17.766  &  17.759  &  17.749  \\
                         &         2                & 17.473  &  17.298  &  16.691  & &  17.579  &  17.542  &  17.455  \\
                         &         3                & 17.274  &  17.062  &  16.109  & &  17.432  &  17.332  &  16.933  \\
0p0h+UCOM+HM             &         4                & 17.212  &  16.988  &  15.980  & &  17.323  &  17.196  &  16.600  \\
                         &         5                & 17.162  &  16.932  &  15.914  & &  17.241  &  17.100  &  16.388  \\
                         &         6                & 17.136  &  16.903  &  15.884  & &  17.196  &  17.047  &  16.293  \\
                         &         7                & 17.124  &  16.890  &  15.871  & &  17.168  &  17.015  &  16.250  \\
                         &         8                & 17.118  &  16.884  &  15.865  & &  17.145  &  16.988  &  16.221  \\\hline
$\Delta_{\textmd{UCOM}}$ &                          & 24.101  &  24.101  &  24.101  & &  23.995  &  23.995  &  23.995  \\
$\Delta_{\textmd{HM}}$   &                          &  0.827  &   1.060  &   2.079  & &   0.682 &    0.831  &   1.585  \\\hline\hline
\end{tabular}%
\end{table*}

Shown in Table \ref{ta03} are the values of the total energy per particle calculated with 0p0h, 0p0h+UCOM, and 0p0h+UCOM+HM wave functions, respectively. The neutron density is taken the normal value $\rho=0.17~\textmd{fm}^{-3}$ as an example. The results with neutron numbers $N=14$ and $N=66$ are both listed here. In the last two lines, the quantities $\Delta_{\textmd{UCOM}}$ and $\Delta_{\textmd{HM}}$ denote the calculated total energy differences between 0p0h and 0p0h+UCOM wave functions and between 0p0h+UCOM and 0p0h+UCOM+HM ($n^{\textmd{max}}_q=6$) wave functions, respectively. The former denotes the contribution of the short-range correlation, while the latter corresponds to the contribution of the high-momentum pairs. It is not surprising we can find from Table \ref{ta03} that for the same neutron number, the short-range correlation contributes the same amount for the three $NN$ potentials. As mentioned above, the differences among the three potentials exist in the channels $^3E$ and $^1O$ which are absent in neutron matter. However, for the two active channels $^1E$ and $^3O$ in neutron matter, the corresponding central parts are identical, which results in the same contribution of the short-range repulsion. Compared to the effect of the short-range correlation, the high-momentum pairs make a relatively smaller contribution to the total energy per particle of neutron matter. For example, with neutron number $N=66$, the short-range repulsion contributes to the total energy all about 23.995 MeV per particle for AV4', AV6', and AV8' $NN$ interactions, while the corresponding attractive effects of high-momentum pairs for the three potentials are only 0.682, 0.831, and 1.585 MeV per particle, respectively. This indicates that for neutron matter the majority correlations coming from the AV4', AV6', and AV8' $NN$ interactions can be treated by the UCOM and the residual part is described by the high-momentum pairs, including tensor correlation and spin-orbit effect. This is consistent to our previous study \cite{ucom4}. Besides, from above values for the high-momentum pairs, we can find that the effect of the high-momentum pairs is successively increased for AV4', AV6', and AV8' $NN$ interactions. This is because of the existence of the additional tensor and spin-orbit forces included in AV6' and AV8' potentials, which will induce more excited high-momentum pairs than those with AV4' potential.

\begin{figure*}[thb]
\centering
\includegraphics[width=1.0\linewidth]{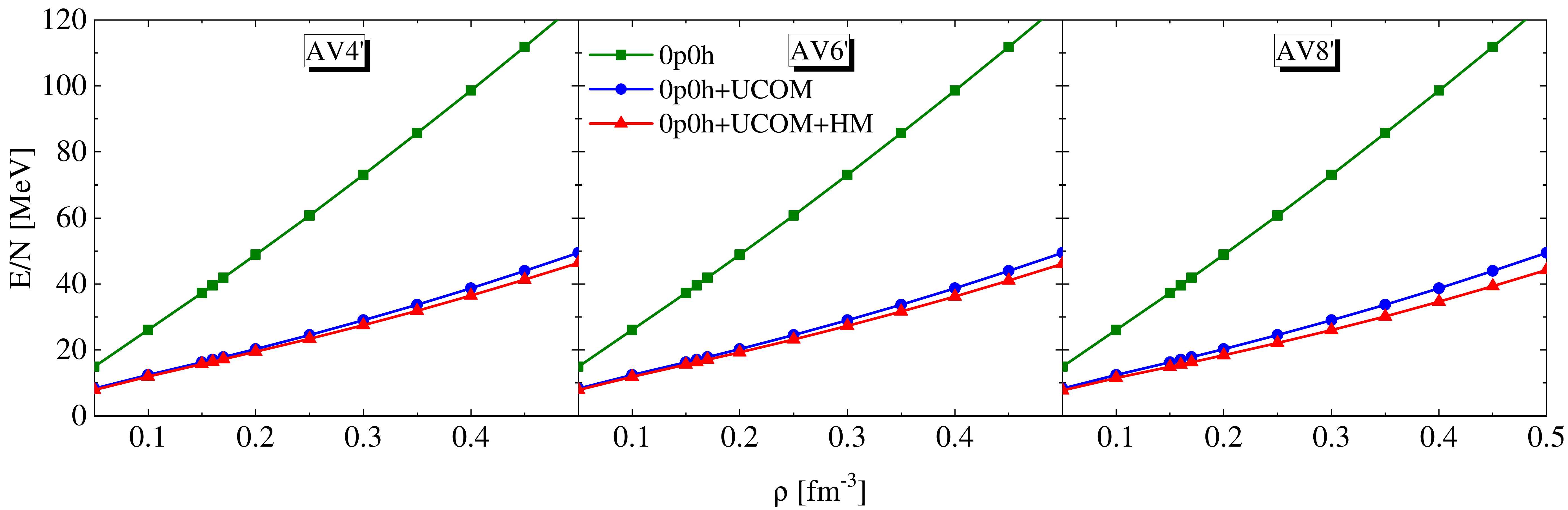}
\caption{Comparison of the EOS for neutron matter calculated with AV4', AV6', and AV8' potentials by using 0p0h, 0p0h+UCOM, and 0p0h+UCOM+HM wave functions.}
\label{figHUph}
\end{figure*}

\begin{figure*}[thb]
\centering
\includegraphics[width=1.0\linewidth]{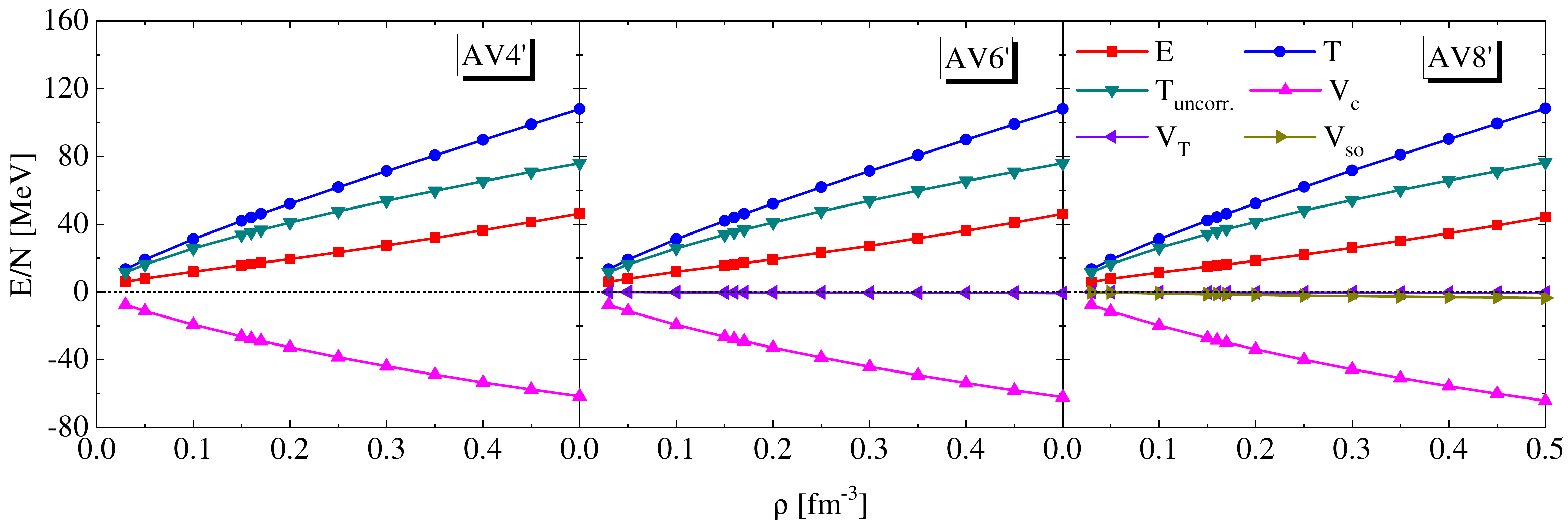}
\caption{Density dependence of all the Hamiltonian components as well as the total energy per particle for neutron matter calculated under the UCOM+HM with AV4', AV6', and AV8' $NN$ interactions. The term $E$ is the total energy, $T$ is the total kinetic energy, $T_{\textmd{uncorr.}}$ is the uncorrelated one-body kinetic energy, and $V_{\textmd{c}}$, $V_{\textmd{T}}$, and $V_{\textmd{so}}$ are the potential energies corresponding to the central, tensor, and spin-orbit forces of the $NN$ interactions, respectively.}
\label{figEOSCOM}
\end{figure*}

\begin{table*}[thb]
\centering
\caption{Values of the Hamiltonian components as well as the total energy per particle for neutron matter at some densities calculated under the UCOM+HM with AV4', AV6', and AV8' $NN$ interactions. Total, kinetic, central part, tensor part, and spin-orbit part of potential energies are given as $E$, $T$, $V_{\textmd{c}}$, $V_{\textmd{T}}$, and $V_{\textmd{so}}$, respectively.}
\label{ta02}
\begin{tabular}{crrrrrrrrrrrrrr}
\hline\hline
\vspace{-4mm}
\kern8mm  &\kern12mm  &\kern12mm  &\kern12mm     &\kern2mm  &\kern12mm  &\kern12mm  &\kern12mm  &\kern12mm  &\kern2mm  &\kern12mm  &\kern12mm  &\kern12mm  &\kern12mm  &\kern12mm \\
$\rho$    &   \multicolumn{3}{c}{AV4'}        &          & \multicolumn{4}{c}{AV6'}       &          &  \multicolumn{5}{c}{AV8'}\\\cline{2-4}\cline{6-9}\cline{11-15}
          &   $E$    &   $T$    & $V_{\textmd{c}}$    &          &   $E$    &   $T$    & $V_{\textmd{c}}$   &  $V_{\textmd{T}}$ & &   $E$    &   $T$    & $V_{\textmd{c}}$    &  $V_{\textmd{T}}$ &  $V_{\textmd{so}}$ \\\hline
0.03  &   5.93 &   13.42 &   -7.49 &  &   5.91 &   13.42 &   -7.47 &   -0.04 &  &   5.87  &  13.43 &   -7.43 &  -0.05 &  -0.09  \\
0.05  &   7.91 &   19.14 &  -11.23 &  &   7.87 &   19.13 &  -11.19 &   -0.08 &  &   7.74  &  19.15 &  -11.07 &  -0.09 &  -0.25  \\
0.10  &  11.97 &   31.23 &  -19.26 &  &  11.88 &   31.23 &  -19.17 &   -0.18 &  &  11.49  &  31.28 &  -18.84 &  -0.18 &  -0.77  \\
0.17  &  17.20 &   46.07 &  -28.87 &  &  17.05 &   46.09 &  -28.75 &   -0.30 &  &  16.29  &  46.22 &  -28.20 &  -0.28 &  -1.45  \\
0.20  &  19.47 &   52.08 &  -32.62 &  &  19.30 &   52.12 &  -32.48 &   -0.34 &  &  18.40  &  52.27 &  -31.85 &  -0.31 &  -1.70  \\
0.30  &  27.52 &   71.35 &  -43.82 &  &  27.30 &   71.41 &  -43.66 &   -0.45 &  &  26.01  &  71.63 &  -42.81 &  -0.40 &  -2.41  \\
0.40  &  36.48 &   89.86 &  -53.38 &  &  36.21 &   89.94 &  -53.19 &   -0.54 &  &  34.62  &  90.23 &  -52.19 &  -0.46 &  -2.96  \\
0.50  &  46.38 &  107.98 &  -61.60 &  &  46.07 &  108.08 &  -61.39 &   -0.61 &  &  44.24  & 108.44 &  -60.27 &  -0.52 &  -3.41  \\\hline\hline
\end{tabular}%
\end{table*}

Compared with the results for neutron number $N=14$ in Table \ref{ta03}, the results for $N=66$ calculated with 0p0h and 0p0h+UCOM wave functions are both a little smaller. By including the high-momentum pairs into the total wave function, i.e., 0p0h+UCOM+HM wave function, the converged total energies with $N=66$ are generally larger than those with $N=14$ at each maximum transfer mode $n^{\textmd{max}}_q$ for the three $NN$ potentials. The trend $E(N=66)>E(N=14)$ is in good agreement with the results given in Ref. \cite{AFDMC2}.

\begin{figure*}[thb]
\centering
\includegraphics[width=1.0\linewidth]{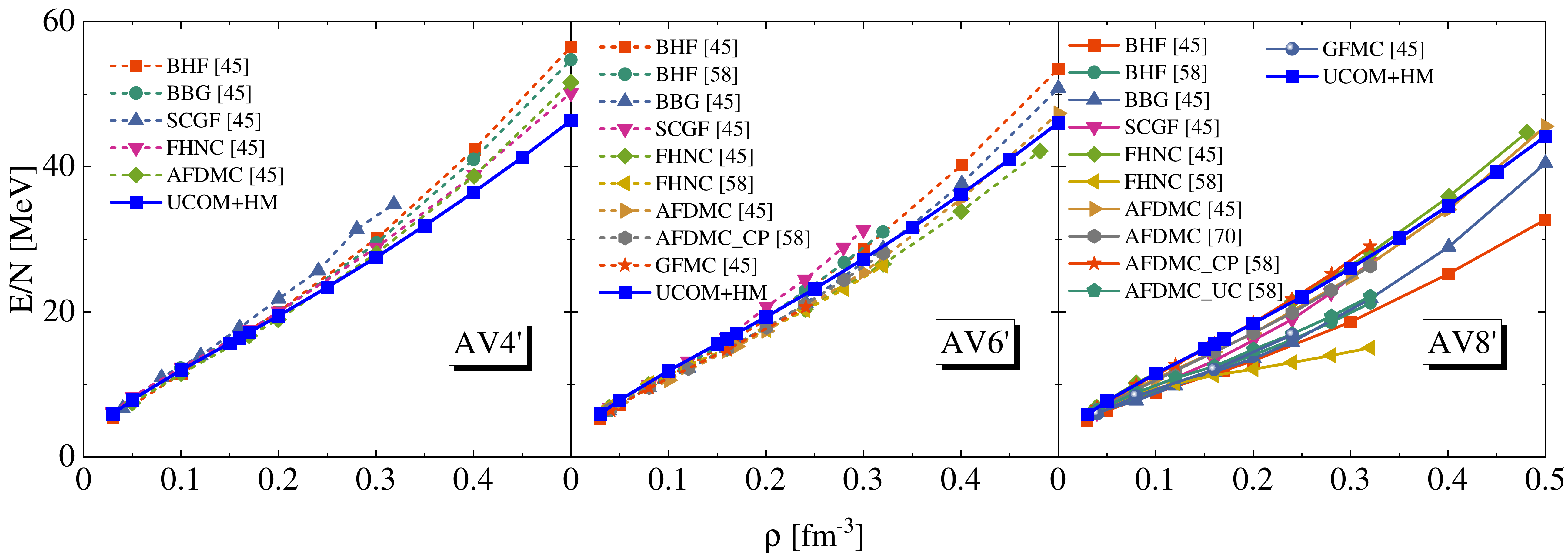}
\caption{Comparison of the EOSs of neutron matter between the UCOM+HM and six other many-body approaches, namely Brueckner-Hartree-Fock (BHF), Brueckner-Bethe-Goldstone (BBG), self-consistent Green's function (SCGF), Fermi hypernetted chain (FHNC), auxiliary field diffusion Monte Carlo (AFDMC), and the Green's function Monte Carlo (GFMC). The results corresponding to AFDMC\_CP and AFDMC\_UC are respectively calculated with constrained- and unconstrained-path approximation for the complex wave functions and propagators to alleviate the sign problem in AFDMC approach \cite{CP1,CP2,CP3}.}
\label{figHMothers}
\end{figure*}

In Fig. \ref{figHUph}, we show the density dependence of the total energy per particle of neutron matter calculated with 0p0h, 0p0h+UCOM, and 0p0h+UCOM+HM wave functions for AV4', AV6', and AV8' $NN$ interactions. It can be immediately noticed from Fig. \ref{figHUph} that for the three $NN$ interactions, the behaviors of the calculated results are very similar by successively using 0p0h, 0p0h+UCOM, and 0p0h+UCOM+HM wave functions. Under only 0p0h wave function, the EOS of neutron matter is rather stiff. After considering the short-range correlation treated by UCOM, i.e., by using 0p0h+UCOM wave function to calculate the EOS, the total energy per particle decreases a lot at all densities. That is to say, the short-range correlation described by UCOM leads to a much more attractive EOS compared with that of 0p0h state without involving any $NN$ correlations. With 0p0h+UCOM+HM wave function including the 2p2h configurations to describe the high-momentum components in neutron matter, the EOS can be further softened, as shown in Fig. \ref{figHUph}.

In Fig. \ref{figEOSCOM}, we show all the Hamiltonian components as well as the total energy per particle dependent on the neutron matter density calculated under the UCOM+HM with AV4', AV6', and AV8' $NN$ interactions. From Fig. \ref{figEOSCOM}, it is found for above three potentials that for neutron matter the kinetic and central part of potential energies mainly contribute to the total energy per particle (absolutely for AV4' potential), while the tensor and spin-orbit parts of potential energy are both relatively small. As a results of the difference between the total and the uncorrelated kinetic energies, the effect of the short-range correlation on the kinetic energy $T-T_{\textmd{uncorr.}}$ is increased with the density of neutron matter for three kinds of $NN$ interactions, which can be as large as 32 MeV per particle with $\rho=0.5~\textmd{fm}^{-3}$. Besides, it can be also seen from the results with AV8' potential that the effects of both the tensor force and the spin-orbit force are definitely small throughout the densities for neutron matter. Though the tensor force is considered to make a large contribution for the symmetric nuclear matter, however, it is very weak for neutron matter due to the absence of the isospin-singlet channels $^3E$ and $^1O$ in neutron matter \cite{Hu1,Hu2,Hu3,BHF1}. The values of the Hamiltonian components as well as the total energy per particle of neutron matter at some densities calculated under the UCOM+HM are summarized in Table \ref{ta02}.

Under the present UCOM+HM framework, we calculate the EOSs of neutron matter by using AV4', AV6', and AV8' $NN$ interactions. In Fig. \ref{figHMothers}, we compare the EOSs provided by UCOM+HM with those calculated under other six different microscopic many-body approaches, namely BHF, BBG, SCGF, FHNC, AFDMC, and GFMC. The results corresponding to AFDMC\_CP and AFDMC\_UC given in Ref. \cite{AFDMC3} are respectively calculated with constrained- and unconstrained-path approximation for the complex wave functions and propagators to alleviate the sign problem in AFDMC approach \cite{CP1,CP2,CP3}. It can be clearly seen from Fig. \ref{figHMothers} that the present results agree with the corresponding ones with above six theories for three $NN$ interactions. In detail, at lower densities the energies calculated by present UCOM+HM are very similar to those of other six theories, while at higher densities there exist differences which increase with the density of neutron matter. In particular, the present results provided by UCOM+HM for the three $NN$ interactions are all very close to those calculated with AFDMC in Refs. \cite{BHF1} and \cite{TBF4} as well as in Ref. \cite{AFDMC3} with the constrained-path approximation. The results corresponding to the unconstrained-path approximation in Ref. \cite{AFDMC3} are much smaller than above four calculations, especially at larger densities, which indicates the importance of the treatment for the sign problem in AFDMC approach. Besides, by comparing the results under UCOM+HM calculated with AV4' and AV6' potentials, it is found that when the tensor force is included into the $NN$ interaction, the difference of the EOS is very small. This is due to the absence of the channels $^3E$ and $^1O$ in neutron matter, especially the former providing the strong tensor correlation. Hence the tensor force has very a few contributions to the total energy per particle of neutron matter, as shown in Table \ref{ta02}. When the spin-orbit force is considered, the present UCOM+HM provides a more attractive EOS in the overall density, by comparing the results of AV8' potential with those of AV4' and AV6' potentials. In present study, we focus on the nuclear EOSs of neutron matter calculated with AV4', AV6', and AV8' $NN$ interactions and study the tensor correlation and spin-orbit effect in neutron matter. In the future, we will employ AV6' and AV8' potentials to further investigate the properties of the symmetric nuclear matter, in which the tensor correlation can play a significant role.

\section{Summary}
\label{iv4}
With our recently proposed variational approach UCOM+HM for describing nuclear matter, we calculate the equation of state (EOS) of neutron matter with AV4', AV6', and AV8' $NN$ interactions, respectively. The neutron matter is described under a finite particle number approach with neutron magic number $N=66$ under a periodic boundary condition. The unitary correlation operator method (UCOM) is used to treat the short-range correlation induced by the short-range repulsion in the $NN$ interaction. In addition, the 2p2h excitations as high-momentum nucleon pairs (HM) are included to describe the high-momentum components of neutron matter, where the two nucleons in a 2p2h configuration involve a large relative momentum. The 2p2h excitations contribute to treat the non-central tensor and spin-orbit forces in neutron matter.

Under the present UCOM+HM framework, the total energy per particle of neutron matter for above three potentials are all well converged with the increasing 2p2h configurations. By comparing the total energies for neutron matter calculated with 0p0h, 0p0h+UCOM, and 0p0h+UCOM+HM wave functions, the same conclusion is found that for neutron matter the majority correlations coming from the $NN$ interaction can be treated by the UCOM and the residual part is described by the high-momentum pairs, such as tensor correlation and spin-orbit effect. Besides, since the tensor and spin-orbit forces can induce additional excited high-momentum pairs, the effect of the high-momentum pairs is successively increased for AV4', AV6', and AV8' potentials.

We also obtain the density dependence of all the Hamiltonian components as well as the total energy per particle for above three potentials. From the results we find that the kinetic and central part of potential energies mainly contribute to the total energy per particle for neutron matter, while both the tensor and spin-orbit parts of potential energy are relatively small. Besides, the effect of the short-range correlation is found to be increased with the density of neutron matter.

The obtained EOSs of neutron matter calculated under the present UCOM+HM framework are also compared with those of other microscopic many-body theories with the same bare interactions. The comparison shows that the calculated total energy per particle of neutron matter is very similar to those of other approaches, especially consistent to that with AFMDC. Due to the absence of the isospin-singlet channels $^3E$ and $^1O$ in neutron matter, the effect of the tensor correlation therein is found to be very small. When the spin-orbit effect is included into the calculations, a more attractive EOS of neutron matter is obtained.

In the next step, we will extend the investigations to the properties of the symmetric nuclear matter to study both the tensor correlation and the spin-orbit effect, because of the significance of the tensor correlation therein. It is also a very interesting subject to study the effect of three-body force on the EOSs of both neutron and symmetric nuclear matters in the future.

\section*{ACKNOWLEDGEMENTS}
This work was supported by the National Natural Science Foundation of China (Grants No. 11822503, No. 11575082, and No. 11947220), by the Fundamental Research Funds for the Central Universities (Nanjing University), by JSPS KAKENHI Grants No. JP18K03660 and No. JP16K05351, and by a Project funded by China Postdoctoral Science Foundation (Grant No. 2019M661785). The author N. W. would like to thank the support from the foreign young research support program in RCNP, Osaka University.


\begin{thebibliography}{90}

\bibitem{JML}J. M. Lattimer and M. Prakash, Phys. Rep., \textbf{333}: 121 (2000)

\bibitem{CJHo}C. J. Horowitz and J. Piekarewicz, Phys. Rev. Lett., \textbf{86}: 5647 (2001)

\bibitem{AWS}A. W. Steiner, M. Prakash, J. M. Lattimer, P. J. Ellis, Phys. Rep., \textbf{411}: 325 (2005)

\bibitem{BALi}B. A. Li, L. W. Chen, and C. M. Ko, Phys. Rep., \textbf{464}: 113 (2008)

\bibitem{BKSh}B. K. Sharma and S. Pal, Phys. Lett. B, \textbf{682}: 23 (2009)

\bibitem{LWCh}L. W. Chen, C. M. Ko, and B. A. Li, Phys. Rev. Lett., \textbf{94}: 032701 (2005)

\bibitem{DVS}D. V. Shetty, S. J. Yennello, and G. A. Souliotis, Phys. Rev. C, \textbf{76}: 024606 (2007)

\bibitem{BSP}B. S. Pudliner, A. Smerzi, J. Carlson, V. R. Pandharipande, S. C. Pieper, and D. G. Ravenhall, Phys. Rev. Lett., \textbf{76}: 2416 (1996)

\bibitem{PDan}P. Danielewicz and J. Lee, Nucl. Phys. A, \textbf{818}: 36 (2009)

\bibitem{SGan}S. Gandolfi, J. Carlson, and S. C. Pieper, Phys. Rev. Lett., \textbf{106}: 012501 (2011)

\bibitem{SHSh}S. H. Shen, H. Z. Liang, J. Meng, P. Ring, and S. Q. Zhang, Phys. Lett. B, \textbf{778}: 344 (2018)

\bibitem{LWCh2}L. W. Chen, C. M. Ko, and B. A. Li, Phys. Rev. C, \textbf{72}: 064309 (2005)

\bibitem{BGT}B. G. Todd-Rutel and J. Piekarewicz, Phys. Rev. Lett., \textbf{95}: 122501 (2005)

\bibitem{MAF}M. A. Famiano, T. Liu, W. G. Lynch, M. Mocko, A. M. Rogers, M. B. Tsang, M. S. Wallace, R. J. Charity, S. Komarov, D. G. Sarantites, L. G. Sobotka, and G. Verde, Phys. Rev. Lett., \textbf{97}: 052701 (2006)

\bibitem{CXu2}C. Xu and B. A. Li, Phys. Rev. C, \textbf{81}: 044603 (2010)

\bibitem{CXu}C. Xu, Z. Ren, and J. Liu, Phys. Rev. C, \textbf{90}: 064310 (2014)

\bibitem{NWa}N. Wang, M. Liu, X. Z. Wu, and J. Meng, Phys. Lett. B, \textbf{734}: 215 (2014)

\bibitem{Wan}N. Wan, C. Xu, Z. Ren, and J. Liu, Phys. Rev. C, \textbf{97}: 051302(R) (2018)

\bibitem{Liu}Y. Ma, C. Su, J. Liu, Z. Ren, C. Xu, and Y. Gao, Phys. Rev. C, \textbf{101}: 014304 (2020)

\bibitem{Wir1}R. B. Wiringa, V. G. J. Stoks, and R. Schiavilla, Phys. Rev. C, \textbf{51}: 38 (1995)

\bibitem{Piep}S. C. Pieper and R. B. Wiringa, Annu. Rev. Nucl. Part. Sci., \textbf{51}: 53 (2001)

\bibitem{Wir2}R. B. Wiringa and S. C. Pieper, Phys. Rev. Lett., \textbf{89}: 182501 (2002)

\bibitem{Myo11}T. Myo, H. Toki, K. Ikeda, H. Horiuchi, and T. Suhara, Prog. Theor. Exp. Phys., \textbf{2015}: 073D02 (2015)

\bibitem{Myo12}T. Myo, H. Toki, K. Ikeda, H. Horiuchi, and T. Suhara, Phys. Lett. B, \textbf{769}: 213 (2017)

\bibitem{Myo13}T. Myo, H. Toki, K. Ikeda, H. Horiuchi, and T. Suhara, Phys. Rev. C, \textbf{95}: 044314 (2017)

\bibitem{Myo14}T. Myo, H. Toki, K. Ikeda, H. Horiuchi, and T. Suhara, Phys. Rev. C, \textbf{96}: 034309 (2017)

\bibitem{Myo15}T. Myo, H. Toki, K. Ikeda, H. Horiuchi, and T. Suhara, Prog. Theor. Exp. Phys., \textbf{2017}: 073D01 (2017)

\bibitem{Enyo}Y. Kanada-Enyo, M. Kimura, and H. Horiuchi, C. R. Phys., \textbf{4}: 497 (2003)

\bibitem{Yama1}T. Yamada, Ann. Phys., \textbf{403}: 1 (2019)

\bibitem{Yama2}T. Yamada, T. Myo, H. Horiuchi, and H. Toki, arXiv:1808.08120 [nucl-th] (to be published in Prog. Theor. Exp. Phys.).

\bibitem{ucom1}H. Feldmeier, T. Neff, R. Roth, and J. Schnack, Nucl. Phys. A, \textbf{632}: 61 (1998)

\bibitem{ucom2}T. Neff and H. Feldmeier, Nucl. Phys. A, \textbf{713}: 311 (2003)

\bibitem{ucom3}T. Myo, H. Toki, and K. Ikeda, Prog. Theor. Phys., \textbf{121}: 511 (2009)

\bibitem{Myo1}T. Myo, A. Umeya, H. Toki, and K. Ikeda, Phys. Rev. C, \textbf{84}: 034315 (2011); \textbf{86}: 024318 (2012)

\bibitem{Myo2}T. Myo, A. Umeya, K. Horii, H. Toki, and K. Ikeda, Prog. Theor. Exp. Phys., \textbf{2014}: 033D01 (2014)

\bibitem{Hu1}J. Hu, H. Toki, Wu Wen, and H. Shen, Phys. Lett. B, \textbf{687}: 271 (2010)

\bibitem{Hu2}J. Hu, H. Toki, Wu Wen, and H. Shen, J. Basic Appl. Phys., \textbf{1}: 1 (2012)

\bibitem{Hu3}J. Hu, H. Toki, and Y. Ogawa, Prog. Theor. Exp. Phys., \textbf{2013}: 103D02 (2013)

\bibitem{Myo3}T. Myo, Prog. Theor. Exp. Phys., \textbf{2018}: 031D01 (2018)

\bibitem{Lyu1}M. Lyu, M. Isaka, T. Myo, H. Toki, K. Ikeda, H. Horiuchi, T. Suhara, and T. Yamada, Prog. Theor. Exp. Phys., \textbf{2018}: 011D01 (2018)

\bibitem{Lyu2}M. Lyu, T. Myo, M. Isaka, H. Toki, K. Ikeda, H. Horiuchi, T. Suhara, and T. Yamada, Phys. Rev. C, \textbf{98}: 064002 (2018)

\bibitem{Myo4}T. Myo, H. Toki, K. Ikeda, H. Horiuchi, T. Suhara, M. Lyu, M. Isaka, and T. Yamada, Prog. Theor. Exp. Phys., \textbf{2017}: 111D01 (2017)

\bibitem{Zhao}Q. Zhao, M. Lyu, Z. Ren, T. Myo, H. Toki, K. Ikeda, H. Horiuchi, M. Isaka, and T. Yamada, Phys. Rev. C, \textbf{99}: 034311 (2019)

\bibitem{Itag}N. Itagaki and A. Tohsaki, Phys. Rev. C, \textbf{97}: 014304 (2018)

\bibitem{BHF1}M. Baldo, A. Polls, A. Rios, H. J. Schulze, and I. Vida\~{n}a, Phys. Rev. C, \textbf{86}: 064001 (2012) and the references therein

\bibitem{BHF2}H. Q. Song, M. Baldo, G. Giansiracusa, and U. Lombardo, Phys. Rev. Lett., \textbf{81}: 1584 (1998)

\bibitem{BHF3}M. Baldo, G. Giansiracusa, U. Lombardo, and H. Q. Song, Phys. Lett. B, \textbf{473}: 1 (2000)

\bibitem{SCGF1}W. H. Dickhoff and C. Barbieri, Prog. Part. Nucl. Phys., \textbf{52}: 377 (2004)

\bibitem{SCGF2}A. Rios, A. Polls, and I. Vida\~{n}a, Phys. Rev. C, \textbf{79}: 025802 (2009)

\bibitem{SCGF3}V. Som\`{a} and P. Boz\.{e}k, Phys. Rev. C, \textbf{74}: 045809 (2006); \textbf{78}: 054003 (2008)

\bibitem{SCGF4}A. Rios, A. Polls, and W. H. Dickhoff, Phys. Rev. C, \textbf{79}: 064308 (2009)

\bibitem{FHNC1}V. R. Pandharipande and R. B. Wiringa, Rev. Mod. Phys., \textbf{51}: 821 (1979)

\bibitem{FHNC2}R. B. Wiringa, V. Fiks, and A. Fabrocini, \emph{ibid}, \textbf{38}: 1010 (1988)

\bibitem{FHNC3}S. Fantoni and A. Fabrocini, \emph{Microscopic Quantum ManyBody Theories and Their Applications}, Lecture Notes in Physics Vol. 510 (Springer, Berlin, 1998)

\bibitem{FHNC4}A. Lovato, O. Benhar, S. Fantoni, A. Y. Illarionov, and K. E. Schmidt, Phys. Rev. C, \textbf{83}: 054003 (2011)

\bibitem{AFDMC1}A. Sarsa, S. Fantoni, K. E. Schmidt, and F. Pederiva, Phys. Rev. C, \textbf{68}: 024308 (2003)

\bibitem{AFDMC2}S. Gandolfi, A. Y. Illarionov, K. E. Schmidt, F. Pederiva, and S. Fantoni, Phys. Rev. C, \textbf{79}: 054005 (2009)

\bibitem{AFDMC3}M. Piarulli, I. Bombaci, D. Logoteta, A. Lovato, and R. B. Wiringa, Phys. Rev. C, \textbf{101}: 045801 (2020)

\bibitem{GFMC1}J. Carlson, J. J. Morales, V. R. Pandharipande, and D. G. Ravenhall, Phys. Rev. C, \textbf{68}: 025802 (2003)

\bibitem{GFMC2}J. Carlson, S. Gandolfi, F. Pederiva, S. C. Pieper, R. Schiavilla, K. E. Schmidt, and R. B. Wiringa, Rev. Mod. Phys., \textbf{87}: 1067 (2015)

\bibitem{CC1}G. Baardsen, A. Ekstr\"{o}m, G. Hagen, and M. Hjorth-Jensen, Phys. Rev. C, \textbf{88}: 054312 (2013)

\bibitem{CC2}G. Hagen, T. Papenbrock, A. Ekstr\"{o}m, K. A. Wendt, G. Baardsen, S. Gandolfi, M. Hjorth-Jensen, and C. J. Horowitz, Phys. Rev. C, \textbf{89}, 014319 (2014).

\bibitem{CC3}J. Lietz, S. Novario, G. R. Jansen, G. Hagen, and M. Hjorth-Jensen, \emph{An Advanced Course in Computational Nuclear Physics: Bridging the Scales from Quarks to Neutron Stars}, Lecture Notes in Physics Vol. 936 (Springer, Berlin, 2017)

\bibitem{QMC1}J. Boronat and J. Casulleras, Phys. Rev. B, \textbf{49}: 8920 (1994)

\bibitem{QMC2}J. Casulleras and J. Boronat, Phys. Rev. Lett., \textbf{84}: 3121 (2000)

\bibitem{TBF0}W. Zuo, A. Lejeune, U. Lombardo, and J. F. Mathiot, Eur. Phys. J. A, \textbf{14}: 469 (2002)

\bibitem{TBF1}K. Hebeler and A. Schwenk, Phys. Rev. C, \textbf{82}: 014314 (2010)

\bibitem{TBF2}S. Gandolfi, J. Carlson, and S. Reddy, Phys. Rev. C, \textbf{85}: 032801 (2012)

\bibitem{TBF3}I. Tews, T. Kr\"{u}ger, K. Hebeler, and A. Schwenk, Phys. Rev. Lett., \textbf{110}: 032504 (2013)

\bibitem{TBF4}S. Gandolfi, J. Carlson, S. Reddy, A. W. Steiner, and R. B. Wiringa, Eur. Phys. J. A, \textbf{50}: 10 (2014)

\bibitem{TBF5}C. Drischler, A. Carbone, K. Hebeler, and A. Schwenk, Phys. Rev. C, \textbf{94}: 054307 (2016)

\bibitem{ucom4}T. Myo, H. Takemoto, M. Lyu, N. Wan, C. Xu, H. Toki, H. Horiuchi, T. Yamada, and K. Ikeda, Phys. Rev. C, \textbf{99}: 024312 (2019)

\bibitem{Lin}C. Lin, F. H. Zong, and D. M. Ceperley, Phys. Rev. E, \textbf{64}: 016702 (2001)

\bibitem{Power1}D. A. Mazziotti, Phys. Rev. A, \textbf{57}: 4219 (1998)

\bibitem{Power2}D. C. Lay, S. R. Lay, and J. J. McDonald, \emph{Linear Algebra and Its Applications}, Pearson Education, 5th ed. (2015)

\bibitem{Power3}F. Colmenero and C. Valdemoro, Int. J. Quantum Chem., \textbf{51}: 369 (1994)

\bibitem{CP1}K. E. Schmidt and M. H. Kalos, \emph{Applications of the Monte Carlo Method in Statistical Physics}, edited by K. Binder, Springer-Verlag, Heidelberg, (1984)

\bibitem{CP2}S. Zhang, J. Carlson, and J. E. Gubernatis, Phys. Rev. B, \textbf{55}: 7464 (1997)

\bibitem{CP3}S. Zhang and H. Krakauer, Phys. Rev. Lett., \textbf{90}: 136401 (2003)


\end{thebibliography}
\end{document}